\documentclass[12pt,
tightenlines,
floats,
showpacs,
nofootinbib,
amsmath,
amssymb,
aps,
superscriptaddress,
prd]{revtex4-1}

\usepackage{slashed}
\usepackage{textcomp}
\usepackage{mathrsfs} 
\usepackage{amsmath,amssymb}

\def\R{\mathbb{R}}	
\def\be{\begin{equation}}
\def\ee{\end{equation}}
\def\rep{representation}
\def\reps{representations}
\def\quan{quantization}	
\def\U{{\cal U}}	
\def\V{{\cal V}}	
\def\H{{\cal H}}	

\def\W{{\cal W}}
\def\S{{\cal S}}

\begin{document}

\title{A Brief Overview of Results about \\ Uniqueness of the Quantization in Cosmology}

\author{Jer\'onimo Cortez}
\email{jacq@ciencias.unam.mx}
\affiliation{Departamento de F\'isica, Facultad de Ciencias, Universidad Nacional Aut\'onoma de M\'exico, \mbox{Ciudad de M\'exico~04510, M\'exico}}
\author{Guillermo A.~Mena Marug\'an}
\email{mena@iem.cfmac.csic.es}
\affiliation{Instituto de Estructura de la Materia, IEM-CSIC, Serrano 121, 28006~Madrid, Spain}
\author{Jos\'e M.~Velhinho}
\email{jvelhi@ubi.pt}
\affiliation{Faculdade de Ci\^encias and FibEnTech-UBI, Universidade da Beira Interior, R. Marqu\^es D'\'Avila e Bolama, 6201-001 Covilh\~a, Portugal}

\begin{abstract}The purpose of this review is to provide a brief overview of some recent conceptual developments about possible criteria to guarantee the uniqueness of the quantization in a variety of situations that are found in cosmological systems. These criteria impose some conditions on the representation of a group of physically relevant linear transformations. Generally, this group contains any existing symmetry of the spatial sections. These symmetries may or may not be
sufficient for the purpose of uniqueness and may have to be complemented with other remaining symmetries that affect the time direction, or with dynamical transformations that in fact are not symmetries. We discuss the extent to which a unitary implementation of the resulting group suffices to fix the quantization, a demand that can be seen as a weaker version of the requirement of invariance. In particular, a strict invariance under certain transfomations may eliminate some physically interesting possibilities in the passage to the quantum theory. This is the first review in which this unified perspective is adopted to discuss otherwise rather different uniqueness criteria proposed either in homogeneous loop quantum cosmology or in the Fock quantization of inhomogeneous cosmologies.
\end{abstract}

\maketitle

\section{Introduction}
\label{Intro}

Quantization is the process of constructing a description that incorporates the principles of Quantum Mechanics starting with a given classical system. 
In the present work we consider exclusively the so-called canonical quantization process. This means that the classical system can be described in canonical form (e.g. in terms of variables that form canonical pairs)
and that one aims at promoting classical variables to operators in a Hilbert space, preserving the canonical structure as much as possible.  The prototypical system is, of course, the phase space  ${\mathbb{R}}^{2n}$, with coordinates $\{(q_i,p_i), i=1, \cdots,n\}$, equipped  with the Poisson bracket $\{q_i,p_j\}=\delta_{ij}$. The standard quantization is realized in the Hilbert space $L^2(R^n)$ of square integrable functions, with configuration variables promoted to multiplicative operators 
\begin{equation}
\label{2}
\hat{q}_i\psi=q_i{\psi},
\end{equation} 
and momentum variables acting as derivative operators
\begin{equation}
\label{1}
\hat{p}_i\psi=-i \frac{\partial}{\partial q_i}\psi.
\end{equation} 
Here, and in the following, we set the reduced Planck constant $\hbar$ equal to one. These operators satisfy the canonical commutation relations (CCRs), thus implementing Dirac's quantization rule that {\em Poisson brackets go to commutators} \cite{dirac}. This quantization is irreducible, in the sense that any operator that commutes with all the $\hat{q}_i$'s and $\hat{p}_i$'s is necessarily proportional to the identity operator. In addition, this quantization satisfies a technical continuity condition, since it provides a continuous representation of the Weyl relations associated with the CCRs (see Section \ref{weyl} for details). It turns out that these two conditions uniquely determine the quantization. This is the celebrated Stone--von Neumann uniqueness theorem (see e.g.~Ref.~\cite{RS1}): every irreducible representation of the CCRs coming from a continuous representation of the Weyl relations is unitarily equivalent to the aforementioned quantization, meaning that, given operators  $\hat{q}'_i$ and $\hat{p}'_i$ with the above properties in a Hilbert space $\cal H$, there exists a unitary operator ${U}:{\cal H}\to L^2(R^n)$ relating the two sets of operators $\hat{q}_i$, $\hat{p}_i$ and $\hat{q}'_i$, $\hat{p}'_i$ (see Equation (\ref{7}) below).

On the other hand, it is known since Dirac's work (and it was rigorously proved by Groenewold and van Hove \cite{gr,vh}) that imposing Dirac's quantization rule on a large set of observables is not viable, in the sense that it is impossible to satisfy the relations 
\begin{equation}
\label{4}
[\hat{f}, \hat{g}]= i\widehat{\{f,g\}}
\end{equation} 
for a large set of classical observables, and certainly not for the whole algebra of classical observables (see Ref.~\cite{gotay} for a thorough discussion).

Nevertheless, in any given physical system there are certainly observables of interest, other than the coordinate variables $\hat{q}_i$ and $\hat{p}_i$, which need to be quantized as well. Also, certain canonical transformations typically stand out in a given classical system, e.g. dynamics or symmetries, and these require as well a proper quantum treatment (these two aspects are in fact related, since observables of physical interest often emerge as generators of some special groups of canonical transformations). This poses no problem as far as one considers the standard quantization of linear canonical transformations and the corresponding generators in ${\mathbb{R}}^{2n}$, precisely because of the above uniqueness theorem. One can easily illustrate this issue by considering  a canonical transformation
\begin{align}
\label{5}
\begin{pmatrix}
q \\ p
\end{pmatrix}\to\begin{pmatrix}
q' \\ p' 
\end{pmatrix}=S\begin{pmatrix}
q \\ p
\end{pmatrix}
\end{align}
in ${\mathbb{R}}^{2n}$, where $S$ is a symplectic matrix. Then, the operators $\hat{q}'_i$ and $\hat{p}'_i$, defined as
\begin{align}
\label{6}
\begin{pmatrix}
\hat{q}' \\ \hat{p}'\end{pmatrix}
=S\begin{pmatrix}
\hat{q} \\ \hat{p}
\end{pmatrix},
\end{align}
provide a new representation of the CCRs (in this case in the same Hilbert space) with the same properties of irreducibility and continuity. So, it is guaranteed that there exists a unitary operator $U$ such that
\begin{equation}
\label{7}
\hat{q}'_i={U}^{-1}\hat{q}_i{U}, \quad \hat{p}'_i={U}^{-1}\hat{p}_i{U}.
\end{equation} 
The unitary operator ${U}$ is naturally interpreted as the quantization of the symplectic transformation $S$. We will also refer to it as the unitary implementation (at the quantum level) of the canonical transformation $S$. If instead of a single linear transformation $S$, one has a 1-parameter group, generated e.g. by a quadratic Hamiltonian function $H$, one obtains a corresponding 1-parameter group of unitary operators ${U}(t)$, which is typically continuous, so that a self-adjoint generator $\hat H$ such that ${U}(t)=e^{-i\hat{H}t}$ can be extracted. Non-quadratic classical Hamiltonians of the type $H=(1/2)\sum p_i^2 + V(q_i)$ do not fall in this last category, but there is nevertheless a standard and well defined procedure to obtain their quantum version (see Ref. \cite{RS} for details), which is simply to define $\hat H$ as $\hat H =(1/2)\sum \hat{p}^2_i +V(\hat{q}_i)$.

Some ambiguities may occur in the quantization of more general functions, involving e.g. products of $q$'s and $p$'s, but these ambiguities are typically not too severe. These are precisely the type of ambiguities that may happen in standard homogeneous quantum cosmology (QC).  In fact, in a homogeneous cosmological model, the number of both gravitational and matter degrees of freedom (DoF) is hugely reduced, ending up with just a finite number of global DoF, precisely due to homogeneity. In this so-called minisuperspace setup, the configuration variables on the gravitational side are typically given by the different scale factors, the number of which depend on the degree of anisotropy. The object of interest here is the Hamiltonian constraint, the quantization of which leads to what is often called the Wheeler--de Witt equation $\hat{H}\Psi=0$. This quantization may not be entirely trivial, owing to a possibly complicated dependence of the constraint with respect to the basic configuration and momentum variables. Nevertheless, the ambiguity that may emerge from the quantization of the Hamiltonian constraint typically involves a choice of factor ordering in $\hat H$ with respect to the basic quantum operators. Although different choices may lead to different versions of the Hamiltonian constraint, it is often the case that this does not affect the physical predictions substantially, in the sense that the predictions remain qualitatively the same.

Thus, the formalism of standard homogeneous QC, based on the standard quantization for a finite number of DoF, is to a large extent free of major ambiguities, as follows from the Stone--von Neumann theorem.

This last paradigm can be broken in two different ways, from very distinct reasons. First, the quantization -- even of the set of basic configuration and momentum variables -- of systems with an infinite number of DoF escapes the conclusions of the Stone--von Neumann theorem. On the contrary, in that case there are representations of the CCRs leading to physically inequivalent descriptions of the same system. This occurs when local DoF are considered, of which one can mention two distinct situations of interest in cosmology: i) the quantization of gravitational DoF (and possibly also of matter fields) in inhomogeneous cosmologies (such as in the example of Gowdy models \cite{Gowdy}), and ii) the quantum treatment of fields propagating in a non-stationary curved spacetime (e.g. of the FRW or de Sitter type) which is considered as a classical background. While the first case embodies genuine applications to QC, i.e. a full quantum treatment of the gravitational DoF (in cases with considerable symmetry, like the Gowdy model, in the so-called midisuperspace setup), the second situation finds applications in the treatment of quantum perturbations in cosmology (both of gravitational and matter DoF), with the homogeneous background kept as a classical entity, like e.g. in inflationary scenarios. There is thus the need of selecting physically relevant quantizations corresponding to a given system containing an infinite number of DoF. Note that available selection criteria (leading to uniqueness) typically rely on stationarity, and are therefore not applicable in the above described situations. 

Precisely, Section \ref{Sec:Unit-DF} is devoted to review selection criteria that were recently introduced and proved viable, leading to unique and well defined quantizations in the aforementioned cases. Such criteria are based on remaining symmetries, present in the cosmological system, and crucially on the unitary implementation of the dynamics, which can be seen as a weaker version of the requirement of invariance under time-translations, which can be applied only in stationary settings. 

The other avenue for departure from the conclusions of the  Stone--von Neumann theorem which is relevant in the cosmological context is exemplified by the quantization approach for homogeneous cosmologies known as loop quantum cosmology (LQC)\footnote{There are nowadays also LQC-inspired applications to inhomogeneous cosmologies. We will not consider them in the present work.}. Although  the same models with a finite number of DoF are considered as in standard homogeneous cosmology, the obtained quantizations are not physically equivalent. The type of quantization used in LQC is not unitarily equivalent to that of standard QC, and the reason why this is possible is because in LQC one of the conditions of the Stone--von Neumann theorem is broken, in a hard way. The LQC-type of quantization starts from the Weyl relations, which are the exponentiated version of the CCRs, and considers representations of the Weyl relations that are not continuous, thus violating one of the conditions of the Stone--von Neumann theorem. In particular, it is the configuration part of the representation that is not continuous. In result, and  although  the correspondent of the unitary group $e^{it\hat q}$ is well defined in the LQC-type of representation as a unitary group ${U}(t)$, the would be generator $\hat q$ cannot be defined, owing to the lack of continuity. This, in turn, is at the heart of the emergence of the discretization (in the canonically conjugate variable) that is so characteristic of LQC. Together with quantization methods adapted from those of loop quantum gravity (LQG) \cite{LQG1,LQG2}, this discretization is responsible, at the end of the day, of the results about singularity avoidance for which LQC is known.
One question that naturally arises is the following: are there other representations of the Weyl relations with different physical properties? Or is this particular LQC-type of representation naturally selected in some way? In Section \ref{lqc} we discuss and comment on a uniqueness result for {isotropic LQC recently put forward by Engle, Hanusch, and Thiemann \cite{EHT}, following a previous discussion concerning the Bianchi I case \cite{AC}}.

We would like to stress that, to the best of our knowledge, this is the first time that the results reviewed in Section \ref{lqc} and those mentioned in Section \ref{Sec:Unit-DF} are considered and discussed together. In particular, this joint and integrated review brings about a discussion on the two possible ways to use relevant transformations
in order to select a unique quantization. In fact, results like those described in Section \ref{lqc} are rooted on a requirement of strict invariance, while the results of Section \ref{Sec:Unit-DF} relax that condition (in what dynamical transformations are concerned),  requiring only the weaker condition of unitary implementation of the transformations in question. These two approaches are discussed, providing a better understanding on the results achieved so far in cosmology.

For completeness, we will start with a very brief review of the formalism for the study of Weyl algebras and their representations. Also, we include an appendix sketching the proof of the uniqueness of the representation results mentioned in Section \ref{Sec:Unit-DF}, in the simplest case of a scalar field in $S^1$ with
time-dependent mass, with the purpose of providing the main steps and typical arguments of the proofs to the interested reader, without overloading the main text. 

\section{Weyl Algebra and Standard Representations}
\label{weyl}

Let $\U$ and $\V$ be a pair of unitary \reps\ of the commutative group $\R$, in the same Hilbert space $\cal H$, i.e. $\U(a)$ and $\V(b)$ are unitary operators for all real values of $a$ and $b$ and such that $\U(a+a')= \U(a)\U(a')$ and $\V(b+b')= \V(b)\V(b')$. The pair $\U$, $\V$ is said to satisfy the Weyl relations if
\be
\label{8}
\V(b)\U(a)= e^{i ab}\U(a)\V(b).
\ee 

The standard \rep\ of the Weyl relations, corresponding to the usual Schr\"odinger \rep\ of the CCRs, is obtained as follows. Consider the Hilbert space $L^2(\R)$ of square integrable functions $\psi(q)$ with respect to the usual Lebesgue measure $dq$. The expressions
\be
\label{9}
\left(\U(a)\psi\right)(q)= e^{iaq}\psi(q)
\ee
and 
\be
\label{10}
\left(\V(b)\psi\right)(q)= \psi(q+b)
\ee
define unitary \reps\ of $\R$ which clearly satisfy the Weyl relations (\ref{8}). These \reps\ are moreover jointly irreducible and continuous, i.e. $a\mapsto \U(a)$ and $b\mapsto \V(b)$ are continuous functions. The appropriate notion of continuity of these operator valued  functions is that of strong continuity, and irreducibility means that no proper subspace of $\cal H$ supports the action of both $\U(a)$ and $\V(b)$ $\forall a, b$. 

It is precisely due to continuity that Stone's theorem guarantees that it is possible to define infinitesimal generators $\hat q$ and $\hat  p$ such that  $\U(a)=e^{ia\hat q}$ and $\V(b)=e^{ib\hat p}$. In this case, it turns out that $\hat q$ is the multiplication operator $q$ and $\hat p=-i\frac{d}{dq}$.

The celebrated {Stone--von Neumann Theorem} ensures that any other \rep\ of the Weyl relations on a separable Hilbert space, with the same properties of irreducibility and continuity, is unitarily equivalent to the one above.

In order to make contact with the language of $\star$-algebras, we now introduce the so-called  Weyl algebra. This is the algebra of formal products of objects $\U(a)$ and $\V(b)$, subjected to the Weyl relations (\ref{8}). Note that, thanks to the Weyl relations, a generic element of the Weyl algebra can always be written as a finite linear combination of elements of the form $\U(a)\V(b)$, $a,b\in\R$, or equivalently of elements 
\be
\label{Weyl}
\W(a,b)=e^{i ab/2}\U(a)\V(b),
\ee
known as Weyl operators.

A \rep\ of the Weyl relations is thus  tantamount to a \rep\  of the Weyl algebra, and since this is a $\star$-algebra with identity, its  \reps\  can be discussed in terms of states of the algebra\footnote{A state $\omega$ of a $\star$-algebra $\cal A$ is a linear functional such that $\omega(a a^*)\geq 0$, $\forall a\in\cal A$, and $\omega({\bf 1})=1$, where  ${\bf 1}$ is the identity of the algebra and the symbol $\,^*$ denotes the involution operation, e.g.~complex conjugation in algebras of functions and adjointness in algebras of operators.}. In particular, given a state one can construct  a cyclic \rep\ of the algebra, by means of the so-called {GNS construction \cite{BR}}. Taking into account the above remarks, it follows that states of the Weyl algebra, and therefore the corresponding \reps, are uniquely determined by the values assigned to the Weyl operators.

Concerning the unitary implementation of automorphisms of $\star$-algebras, it is a well known fact that, if a state $\omega$ is invariant under a given transformation, then a unitary implementation of that transformation is ensured to exist in the GNS representation defined by $\omega$. 

The above constructions related to the Weyl algebra are straightforwardly generalized to any finite number of DoF, and also without difficulties to field theories. In this respect, let us consider for instance a scalar field in $\R^3$.

The starting point in the canonical quantization process is the choice of properly defined variables in phase space, that are going to play e.g. the role of the $q$'s and the $p$'s. The integration of the field $\phi$ and of the canonically conjugate momentum $\pi$ against smooth and fast decaying {\em test} functions provides just those amenable variables. Thus, given test functions $f$ and $g$ with the above properties, hence belonging to the so-called Schwartz space $\cal S$, one defines linear functions in phase space by
\be
\label{obs}
(\phi,\pi)\mapsto \int\phi f\, d^3 x + \int\pi g\, d^3 x\, =:\, \phi(f)+\pi(g).
\ee
In particular, the variables $\phi(f)$ and $\pi(g)$, with Poisson bracket
\be
\{\phi(f),\pi(g)\}=\int fg\, d^3 x,
\ee
replace in this context the familiar  $q$'s and $p$'s. The corresponding 
Weyl relations are
\be
{\cal V}(g) {\cal U}(f)= e^{i\int fg d^3 x}{\cal U}(f){\cal V}(g), \quad\quad f,g\in {\cal S}, 
\ee
with seemingly defined Weyl operators
\be
\W(f,g)=e^{\frac{i}{2}\int fgd^3x}\U (f)\V(g).
\ee
Let us focus on a particular type of \reps\ of the Weyl relations (or equivalently of the associated Weyl algebra), namely \reps\ of the Fock type. These are \reps\ defined by complex structures on the phase space, or equivalently on the space $\S\oplus\S$ of pairs of test functions $(f,g)$. Note in this respect that the pairs $(f,g)$ define linear functionals in the phase space and so $\S\oplus\S$ is naturally dual to the phase space, inheriting therefore a symplectic structure which is induced from that originally considered on phase space. We also recall that a complex structure $J$ on a linear space with symplectic form $\Omega$ is a  linear symplectic transformation such that $J^2=-\bf 1$, compatible with $\Omega$ in the sense that the  bilinear form defined by $\Omega(J\cdot,\cdot)$ is positive definite. 

Let then $J$ be a complex structure on the symplectic space $\S\oplus\S$. As mentioned above, a state of the Weyl algebra is defined by the values on the Weyl operators, and therefore the assignment 
\be
\label{fock}
\W(f,g)\mapsto e^{-\frac{1}{4}\Omega\left(J(f,g),(f,g)\right)}
\ee
defines a state and an associated cyclic \rep\ of the Weyl algebra. The cyclic vector is here physically interpreted as the vacuum of the Fock \rep, and therefore the above expression coincides precisely with the expectation values of the Weyl operators on the vacuum  of the Fock \rep\ defined by $J$. 

Let us now discuss the question of unitary implementation of symplectic transformations in the specific context of Fock \reps. We consider unitary operators $\W(f,g)$ providing a \rep\ of the Weyl algebra, and a linear canonical transformation $A$. There is then a new \rep\ $\W_A$, defined (in the same Hilbert space) by $\W_A(f,g):=\W(A^{-1}(f,g))$. If the \rep\ $\W$ is defined by a complex structure $J$,  it follows that $\W_A$ corresponds to a new complex structure $J_A:=A J A^{-1}$. In general, $A$ is not  unitarily implementable, i.e. there is no unitary operator $U_A$ such that
\be
{U_A^{-1}}\W(f,g){U_A}=\W({A^{-1}}(f,g)).
\ee
In fact, the Fock \reps\ defined by $J$ and $J_A$ are unitarily equivalent  if and only if the difference $J_A-J$ is an operator of a special type, namely a Hilbert--Schmidt operator \cite{shale}. Two notorious cases where that condition is automatically satisfied are the following. First, every operator in a finite dimensional linear space is of the Hilbert--Schmidt type, and therefore the unitary implementation of symplectic transformations comes for free in finite dimensional phase spaces, as expected from the Stone--von Neumann theorem. On the other hand, the null operator is always of the Hilbert--Schmidt type, regardless of the dimensionality, and therefore a transformation $A$ that leaves $J$ invariant is always unitarily implementable {\em in the Fock \rep\ defined by $J$}. With respect to previous remarks in our exposition, we note that invariance of $J$ immediately translates into invariance of the associated Fock state defined by (\ref{fock}). Of course, the two situations that we have described correspond only to sufficient conditions for unitary implementation, which are by no means necessary. In particular, unitary implementation of a canonical transformation $A$ can be achieved via a non-invariant complex structure $J$, provided that $J_A-J$ is Hilbert--Schmidt.

In any case, and in clear contrast with the situation found for a finite number of DoF, in field theory no Fock \rep\ supports the unitary implementation of the full group of linear canonical transformations. Fock \reps\ are therefore distinguished by the class of transformations that are unitarily implementable. Now, in a particular theory, specified e.g. by a given Hamiltonian, a particular set of canonical transformations stands out, namely transformations generated by the Hamiltonian and by possible symmetries. The requirement of unitary implementation of relevant canonical transformations therefore provides a criterion guiding the selection of one \rep\ over another when we are trying to quantize a field theory. In this respect, we notice that the case of the set of transformations corresponding to classical time evolution is particularly relevant, given the role that unitarity plays in the probabilistic interpretation of the quantum theory.

The simplest situation is that of a free field of mass $m$ in Minkowski spacetime. In this case the \rep\  is completely fixed by the requirement of invariance under spatial symmetries and time evolution (or the full Poincar\'e group), in the sense that a unique complex structure is selected by that requirement, namely $J_m$ defined by
\be
\label{jm}
J_m(f,g)=\left((m^2-\Delta)^\frac{1}{2}g,-(m^2-\Delta)^{-\frac{1}{2}}f\right),
\ee
where $\Delta$ is the Laplacian.

Besides the above free field in Minkowski spacetime, there are other known situations of linear dynamics where uniqueness results apply. In fact, provided that the Hamiltonian is time independent, the criterion of positivity of the energy, together with invariance under the 1-parameter group of canonical transformations generated by the Hamiltonian, is sufficient to select a unique complex structure. Here, positivity means that the unitary group implementing the dynamics possesses a positive generator, i.e. the quantum Hamiltonian is a positive operator \cite{kay,BSZ}. This result finds remarkable applications in the quantization of free fields in stationary curved spacetimes (i.e. with a timelike Killing vector) \cite{AM,qftwald}. On the other hand, no general uniqueness results are available for the non-stationary situations typical in cosmology.

\section{Loop Quantum Cosmology}
\label{lqc}

Let us now consider the \rep\ of the Weyl relations used in LQC, sometimes referred to as the polymer \rep.  We will restrict our attention to its simplest version, namely the one associated with the homogeneous and isotropic flat FLRW model\footnote{We also restrict attention to the more usual formulation of LQC, leaving aside the so-called Fleischhack approach, which is also considered in Ref. \cite{EHT}.}. For convenience, we set the speed of light and Newton constant multiplied by $4\pi$ equal to the unit, and we make the Immirzi parameter \cite{Imm} equal to 3/2 in order to simplify our equations, without loss of generality. At the classical level, the system is described by a pair of canonically conjugate variables, usually denoted by $c$ and $p$, with Poisson bracket $\{c,p\}=1$. The variables $c$ and $p$ parametrize, respectively, the (homogeneous) Ashtekar connection and the densitized triad (see Ref. \cite{EHT} for details in the context of the current uniqueness discussion, and Refs. \cite{ABL,AL} for more general introductions to LQC). In particular, $p$ is proportional to the square scale factor of the FLRW spacetime.

Let us consider the Hilbert space $\H_{\cal P}$ defined by the discrete measure in $\R$, i.e.~the space of complex functions $\psi(p)$ such that
\be
\label{11}
 \sum_{p\in\R}|\psi(p)|^2<\infty,
\ee
with inner product given by
\be
\label{12}
\langle\psi,\psi^{\prime}\rangle=\sum_{p\in\R}{\bar\psi}(p)\psi^{\prime}(p),
\ee
where the overbar denotes complex conjugation. We note that this Hilbert space, also referred to as the polymer Hilbert space, is very different from the standard one, $L^2(\R)$. In particular, $\H_{\cal P}$ is non-separable\footnote{Nevertheless, applications of LQC are effectively performed on a separable subspace of $\H_{\cal P}$. This can either be seen as a consistency requirement \cite{comments} or as consequence of superselection \cite{APS}, which in any case can be traced back to the fact that the LQC quantum Hamiltonian constraint is a difference operator of constant step.}. An orthonormal basis of  $\H_{\cal P}$ is formed e.g.~by the uncountable set of functions $\Psi_{p_0}$, for all $p_0\in\R$, where
\be
\label{13}
\Psi_{p_0}(p)=\delta_{pp_0},
\ee
with $\delta_{pp_0}$ being the Kronecker delta.

We then define the operators $\U_{\cal P}(a)$ and $\V_{\cal P}(b)$, for $a, b\in\R$, acting on $\H_{\cal P}$ by
\be
\label{14}
\left(\U_{\cal P}(a)\psi\right)(p)= \psi(p-a)
\ee
and 
\be
\label{15}
\left(\V_{\cal P}(b)\psi\right)(p)= e^{ibp}\psi(p).
\ee
It is clear that these operators satisfy the Weyl relations (\ref{8}) and that the \rep\ is irreducible. The map $b\mapsto \V_{\cal P}(b)$ is continuous, so that one can define the infinitesimal generator. We will denote it by $\pi(p)$, and not $\hat p$, to distinguish it from the standard Schr\"odinger \rep\ in $L^2(\R)$. It follows that
\be
\label{16}
\left(\pi(p)\psi\right)(p)=p\psi(p).
\ee
On the other hand, $\U_{\cal P}(a)$ is not continuous. To see this, it suffices to note that, for arbitrarily small $a$, the vector  $\Psi_{p_0}$ is mapped by $\U_{\cal P}(a)$ to an orthogonal one $\Psi_{p_0+a}$. The would be generator of the unitary group $\U_{\cal P}(a)$ cannot therefore be defined. Nevertheless, taking into account the commutator $[\pi(p), \U_{\cal P}(a)]$, one can see that the operators $\U_{\cal P}(a)$ can be regarded as providing a quantization of the classical variables $e^{iac}$. We note that the reality conditions are properly satisfied, since $\U^{\dag}(a)=\U(-a)$, with the dagger denoting the adjoint. Thus, the set of operators  $\U_{\cal P}(a)$, together with $\pi(p)$, provide a quantization, in the usual Dirac's sense, of the Poisson algebra of phase space functions made of finite linear combinations of the functions $p$ and $e^{iac}$, with $a\in\R$ (see Ref. \cite{bohr} for details).

This Poison algebra is of course different from the kinematical algebra usually associated with the linear phase space $\R^2$ with coordinates  $c$ and $p$, which is simply the Heisenberg algebra of linear (non-homogeneous) functions in $c$ and $p$. At the foundation of LQC, there is thus a situation similar to that of LQG: a non-standard choice of basic variables and a \rep\ of the associated algebra which is non-continuous (at least in part of the algebra), thus obstructing the quantization of the connection itself \cite{LQG1}. However we note that, contrary to the situation in LQG, the standard Schr\"odinger quantization gives us a different \rep\ of the very same Poisson algebra, since the variables $e^{iac}$ are trivially quantized in $L^2(\R)$ by multiplication operators, e.g.~by operators $\widehat{e^{iac}}$ such that
\be
\label{17}
\left(\widehat{e^{iac}}\psi\right)(q)=e^{iaq}\psi(q).
\ee
In the case of this Schr\"odinger quantization, the continuity of the \rep\ $\U(a)$ allows us to define the operator $\hat c$ itself, and therefore to extend the quantization of configuration variables to a much larger set of functions $f(c)$, simply by defining $\widehat{f(c)}=f(\hat c)$, whereas in the polymer quantization one is restricted to quantize configuration variables of the type $e^{iac}$ (and linear combinations thereof).

In LQG there is a celebrated result about the uniqueness of the quantization that gives robustness to the loop \rep\ \cite{LOST,Fleis}. In Ref. \cite{EHT}, the authors proved a similar uniqueness result for LQC, that we now discuss. In order to make contact with that work, which uses the language of $\star$-algebras and corresponding states, we first introduce the LQC analogue of the LQG holonomy-flux algebra, which is again denoted in Ref. \cite{EHT} as the quantum holonomy-flux $\star$-algebra ${\mathfrak{U}}$. The LQC $\star$-algebra ${\mathfrak{U}}$ is constructed in Ref. \cite{EHT} as the algebra of formal products of operators corresponding to the variables $p$ and $e^{iac}$, subjected to the conditions coming from the commutator $[\pi(p), \U_{\cal P}(a)]$. However, since we have already introduced the Weyl algebra, it is more natural here to follow an alternative procedure, and identify instead the  $\star$-algebra ${\mathfrak{U}}$ with the Weyl algebra, i.e. the algebra of formal products of objects $\U(a)$ and $\V(b)$, subjected to the Weyl relations (\ref{8}). 

Recall now that the group of spatial diffeomorphisms is a ``gauge'' symmetry in General Relativity (GR). Physical states in quantum gravity should therefore be invariant under the quantum operators representing these diffeomorphisms (or annihilated by the quantum diffemorphism constraint). Thus, a unitary implementation of the group of spatial diffeomorphisms in the quantum Hilbert space is required. Since the LQG holonomy-flux algebra is again a $\star$-algebra with identity, it follows from previous comments that, in order to achieve the required unitary implementation of the group of diffeomorphisms, it is sufficient that the quantization be defined by a diffeomorphism invariant state.
 
The LQG result of uniqueness of the quantization \cite{LOST,Fleis} guarantees, precisely, that there exists a unique diffeomorphism invariant state of the LQG holonomy-flux algebra. Moreover, the GNS representation defined by such a unique invariant state is unitarily equivalent to the LQG \rep\ that was previously known. The analogous result for LQC starts from the observation, explained in detail in Ref. \cite{EHT}, that a residual gauge group still remains, when descending from full GR to homogeneous and isotropic flat models. In fact, although almost all of the diffeomorphism gauge symmetry is automatically fixed, in homogeneous and isotropic flat models one is left with a small gauge group, namely the group of isotropic dilations, acting on phase space as
\be
\label{18}
(p,c)\mapsto(\lambda p, c/\lambda),\quad \quad \lambda\in\R.
\ee
Thus, there is the possibility of exploring this residual gauge symmetry in order to select a state in LQC, very much like in the above mentioned LQG uniqueness result. The authors of Ref. \cite{EHT} indeed succeeded in proving that there is a unique dilation invariant state of the LQC holonomy-flux algebra, for which the associated GNS \rep\ is unitarily equivalent to the polymer \rep\ described above. 

Although the result of Ref. \cite{EHT} is definitely very interesting and rigorous, we argue here that in a certain sense its status is not as strong as that of the LQG uniqueness result. From the physical viewpoint, what is really required is a unitary implementation of the group of interest (may it correspond to dynamics, to symmetries, or to gauge transformations), and not necessarily an invariant state. There are even situations (see e.g. Section \ref{Sec:Unit-DF}) where invariant states are not available and, nevertheless, a unitary implementation of the relevant group exists. It is true that constructing the quantization by means of an invariant state is sufficient to achieve unitary implementation - and it is perhaps the ``Kings way'' of doing it -, but it is by no means necessary. In the present case, the Schr\"odinger \rep, although not (unitarily equivalent to a \rep) defined by an invariant state, carries a unitary implementation of the group of dilations (\ref{18}), which is physically just as good as the one provided by the polymer \rep. The existence of this unitary implementation actually follows from the Stone--von Neumann theorem, but let us show it explicitly. We consider the transformations $U_{\lambda}: L^2(\R)\to L^2(\R)$ defined by 
\be
\label{19}
\psi(q)\stackrel{U_{\lambda}}{\longmapsto} {\lambda}^{1/2}\psi(\lambda q),\quad\quad \lambda\in\R,
\ee
which are clearly unitary $\forall\lambda$, with respect to the standard inner product defined by the measure $dq$. We consider also the standard operator $\hat p=-i\frac{d}{dq}$ and the Schr\"odinger quantization of configuration variables
\be
\label{20}
\left(\widehat{f(c)}\psi\right)(q)=f(q)\psi(q), \quad \quad \psi\in L^2(\R),
\ee
which obviously includes the LQC variables $e^{iac}$. A straightforward computation shows that
\be
\label{21}
\left(U_{\lambda}^{-1}\hat p U_{\lambda}\psi\right)(q)= -i\lambda\frac{d}{dq}\psi(q),
\ee
\be
\label{22}
\left(U_{\lambda}^{-1}\widehat{f(c)} U_{\lambda}\psi\right)(q)= f(q/\lambda)\psi(q),
\ee
or
\be
\label{23}
U_{\lambda}^{-1}\hat p U_{\lambda}=\widehat{\lambda p}, \quad\quad U_{\lambda}^{-1}\widehat{f(c)} U_{\lambda}=
\widehat{f(c/\lambda)},
\ee
which is the announced unitary implementation of the group of dilations (\ref{18}) in the Schr\"odinger \rep.

From this perspective, we then conclude that the physical criterion of a unitary implementation of the residual group of dilations in homogeneous and isotropic flat cosmology does not fully succeed in selecting a unique quantization, since both the polymer and the Schr\"odinger \reps\ are viable from this viewpoint. Only the more mathematically stringent requirement of strict invariance selects a unique state. This is perhaps a reminder about the fact that strict invariance is not an unavoidable requirement, and that the quantization of groups of interest via unitary implementations that are not \emph{necessarily} based on invariant states is worthwhile exploring.

Let us end with a brief comment regarding the analogous uniqueness result in LQG. In that case, the uniqueness is also proved by requiring strict invariance of a state of the holonomy-flux algebra under the action of spatial diffeomorphisms. There is however a key difference with respect to the above described situation in LQC: no other (irreducible) representation is known, of any kind, admitting a unitary implementation of the diffeomorphism group\footnote{See nevertheless the variations on the LQG \rep\ introduced by Koslowski and Sahlmann \cite{kos,sahl,KS}, and the related developments by M. Varadarajan and Campiglia \cite{VC1,VC2}.} and so there is no alternative route that may cast any shadow on the uniqueness. The situation remains however somewhat open, until a stronger uniqueness result is demonstrated that is based exclusively on a unitary implementation of the spatial diffeomorphisms, and not just in strict invariance, or otherwise until a new \rep\ of the LQG algebra admitting a unitary non-invariant implementation of the diffeomorphism group is constructed.

\section{Fock Quantization in Non-Stationary Cosmological Settings}
\label{Sec:Unit-DF}

As we have already mentioned, the Stone--von Neumann uniqueness result fails for infinite number of DoF, and no general result on the uniqueness of the quantization is available for non-stationary situations. This includes, of course, cases of interest in QC. In some of those cases, the underlying theory can be recast in the form of a linear scalar field with a time-dependent mass, propagating in an auxiliary background spacetime which is both static and spatially compact. One such situation is the linearly polarized Gowdy model with the spatial topology of a three-torus, where the gravitational degrees of freedom are encoded by a scalar field on $S^1$, evolving in time precisely as a linear field with a time-dependent mass of the type $m(t)=1/(4t^2)$ \cite{unigowdy2,ccm1}. Other situations of interest include free scalar fields in cosmological scenarios, e.g. propagating in (compact) FLRW or the Sitter spacetimes. In those cases, the non-stationarity is transfered from the background to the field efective mass, by means of a simple transformation.

\subsection{Gowdy Models}
	
Midisuperspace models are symmetry reductions of full GR that retain an infinite number of degrees of freedom. Typically, these are local degrees of freedom, so that they often describe inhomogeneous scenarios. Therefore, these midisuperspace models must face the inherent ambiguity that affects the quantization of fields. One of the simplest inhomogeneous cosmologies obtained with a symmetry reduction is the linearly polarized Gowdy model on the three-torus, $T^3$ \cite{Gowdy}. This model describes vacuum spacetimes with spatial sections of $T^3$-topology containing linearly polarized gravitational waves, with a symmetry group generated by two commuting, spacelike, and hypersurface orthogonal Killing vector fields. In consequence, the local physical degrees of freedom can be parametrized by a scalar field corresponding to those waves, and effectively living in $S^1$. 
	
After a partial gauge fixing, the line element of the linearly polarized Gowdy $T^3$ model can be written as \cite{CM}
\begin{equation}
\label{24}
ds^{2}=e^{{{\gamma}}-\phi/\sqrt{p}}\left(-dt^{2}+d\theta^{2}\right)+e^{-\phi/\sqrt{p}}t^{2}p^{2}d\sigma^{2}+e^{\phi/\sqrt{p}}d\delta^{2},
\end{equation}
where $(\partial/\partial \sigma)^{a}$ and $(\partial/\partial \delta)^{a}$ are the two Killing vector fields. The true dynamical field DoF are encoded in $\phi(\theta,t)$,  where $t >0$ and  $\theta\in S^{1}$. On the other hand, $p$ is a  homogeneous non-dynamical variable, and the field ${{\gamma}}$ {is completely} determined {by} $p$ and $\phi$ as a result of the gauge-fixing process \cite{CM}.  There remains a~global spatial constraint on the system,  giving rise to the symmetry group of (constant) translations in $S^1$. Time evolution is dictated by the field equation
\begin{equation}
\label{phi-eq-gt2}
\ddot{\phi}+\frac{1}{t}\dot{\phi}-\phi^{\prime\prime}=0.
\end{equation}
Here, the prime denotes the derivative with respect to $\theta$ and the dot denotes the time derivative.
	
A complete quantization of the system was obtained in Ref. \cite{Pierri}. Nevertheless, it was soon realized \cite{CCH-GT3,CM} that the classical dynamics could not be implemented as a unitary transformation in such quantization. With the purpose of achieving unitarity, and restoring in particular the standard probabilistic interpretation of quantum physics within the quantum Gowdy model, an alternative quantization was introduced by Corichi, Cortez, and Mena Marug\'an \cite{unigowdy2,ccm1}. A crucial step towards a quantization with unitary dynamics is the following time-dependent transformation, performed at the  level of the classical phase space. Instead of working with the original field $\phi$ and its corresponding conjugate momentum $P_{\phi}$, the authors introduced a new canonical pair $\chi$ and $P_{\chi}$ related to the first one by means of the canonical transformation
\begin{equation}
\label{ct-gowdy-t3}
\chi=\sqrt{t}\phi,\quad P_{\chi}=\frac{1}{\sqrt{t}}\left(P_{\phi}+\frac{\phi}{2}\right),
\end{equation}
taking advantage in this way of the freedom in the scalar field parametrization of the metric of the Gowdy model.
	
The 	evolution of the new canonical pair turns out to be governed by a time-dependent Hamiltonian that, in our system of units, adopts the expression
\be
\label{25}
H_{\chi}=\frac{1}{2}\oint d\theta\left[P^{2}_{\chi}+\chi^{\prime\,2}+\frac{\chi^{2}}{4t^{2}}\right].
\ee
The corresponding Hamiltonian equations are 
\be
\label{26}
\dot{\chi}=P_{\chi},\quad \dot{P}_{\chi}=\chi^{\prime\prime}-\frac{\chi}{4t^{2}},
\ee
that combined give the second-order field equation 
\begin{equation}
\label{phi-eq-gt3}
\ddot{\chi}-\chi^{\prime\prime}+\frac{\chi}{4t^{2}}=0.
\end{equation}

We can therefore view the system as a linear scalar field with a time-dependent mass of the form $m(t)=1/(4t^2)$, evolving in an effective static spacetime with one-dimensional spatial sections with the topology of the circle\footnote{Alternatively, the system can be considered as an axially symmetric field propagating in a static (2+1)-dimensional spacetime with the spatial topology of a two-torus.}. Another relevant aspect of the model is the invariance of the dynamics under (constant) $S^1$-translations:
\be
\label{27}
T_{\alpha}:\theta\mapsto\theta+\alpha, \quad\quad \alpha\in S^1.
\ee
These translations are moreover symmetries generated by a remaining constraint, as we already mentioned.

The quantization of the system put forward by Corichi, Cortez, and Mena Marug\'an in Refs. \cite{unigowdy2,ccm1} starts from the CCRs satisfied by the canonical pair $\chi$ and $P_{\chi}$, or rather by the corresponding Fourier modes. The advantage of using Fourier components, say $\chi_n=({1}/{\sqrt{2\pi}})\oint d\theta e^{-in\theta}\chi(\theta)$, instead of the field $\chi(\theta)$ itself is clear: since the spatial manifold is compact, the set of Fourier modes is discrete and one therefore avoids the issues of dealing with operator valued distributions, like e.g. $\hat{\chi}(\theta)$. The aforementioned quantization is of the standard Fock type, with the following remarkable properties. To begin with, the complex structure on phase space that effectively defines the Fock \rep\ is invariant under $S^1$-translations. Thus, the corresponding state of the Weyl algebra is $S^1$-invariant, leading to a natural unitary implementation of these gauge transformations. Secondly and most importantly, the classical dynamics in phase space defined by Equation (\ref{26}) (i.e. generated by the Hamiltonian (\ref{25})) is unitarily implemented at the quantum level. In other words, let $t_0$ be an arbitrary but fixed initial time, and $S(t,t_0)$ be the linear symplectic transformation corresponding to the classical evolution in phase space from the time $t_0$ to the arbitrary time $t$. Then, to each transformation $S(t,t_0)$ there corresponds a unitary operator $U(t,t_0)$, that intertwines between the quantum operators $\hat{\chi}$ and $\hat P_{\chi}$ defined at the initial time $t_0$ and those obtained from $\hat{\chi}$ and $\hat P_{\chi}$ by application of $S(t,t_0)$, as exemplified in Equations (\ref{6}, \ref{7}).

This last transformation corresponds of course to the usual evolution in the Heisenberg picture, which can always be formally defined, once canonical operators are given at some initial time. The key difference with respect to systems with a finite number of DoF is that, whereas in those cases the relation between the ``initial'' and the evolved operators in principle is always unitary, the existence of such unitary operators is far from being guaranteed in an arbitrary \rep\ of the CCRs for field theory (or generally with an infinite number of DoF).

We note that, although a unitary implementation of all the transformations $S(t,t_0)$ is achieved in the Corichi, Cortez, and Mena Marug\'an \rep, the corresponding state is not invariant under these transformations. In fact, no state exists such that it remains invariant under all the transformations $S(t,t_0)$, $\forall t$.

All in all, the quantization of the linearly polarized Gowdy model proposed in Refs. \cite{unigowdy2,ccm1} is one of the few available examples of a rigorous and fully consistent quantization of an inhomogeneous cosmological model. Nonetheless, the eventual robustness of its physical predictions might be affected by the possible existence of major ambiguities in the quantization process. Fortunately, a quantization with the aforementioned properties is indeed unique, as shown in Refs. \cite{gowdyqft,cmv1}, where the uniqueness result that we now discuss was derived.

A source of ambiguity in the process leading to the Corichi, Cortez, and Mena Marug\'an \quan\ is the choice of \rep\ for the canonical pair $\chi$ and $P_{\chi}$. However, it was shown in Ref. \cite{gowdyqft} that any other Fock \rep\ of the CCRs that {\em i}) is defined by a $S^1$-invariant complex structure (or equivalently by a $S^1$-invariant state of the Weyl algebra) and {\em ii}) allows a unitary implementation of the dynamics defined by Equation (\ref{26}), is unitarily equivalent to the considered \rep\ (and therefore physically indistinguishable). We note that there actually is an infinite number of $S^1$-invariant states, leading to many inequivalent \reps. It is only after the requirement of unitary dynamics that a unique unitary equivalence class of \reps\ is selected. 

Another possible source of ambiguity concerns the choice of the ``preferred'' canonical pair $(\chi,\ P_{\chi})$. In this respect, note that a time-dependent transformation different from Equation (\ref{ct-gowdy-t3}) would lead to classical dynamics that would not reproduce Equation (\ref{26}), and it is in principle conceivable that the new dynamics could be unitarily implemented in a different \rep, thus leading to a distinct quantization. This is however not the case. In fact, it was shown in Ref. \cite{cmv1} that any other canonical transformation of the type (\ref{ct-gowdy-t3}) modifies the equations of motion in such a way as to render impossible the unitary implementation of the dynamics, with respect to any Fock \rep\ defined by a $S^1$-invariant complex structure. It is worth mentioning that the kind of transformations considered in Ref. \cite{cmv1} is restricted by the natural requirements of locality, linearity, and preservation of $S^1$-invariance. In configuration space, these are contact transformations that produce a time-dependent scaling of the field $\phi$ (see Ref. \cite{cmv1} for a detailed discussion). Such scalings can always be completed into a canonical transformation in phase space, of the general form 
\be
\label{28}
(\phi,P_{\phi})\mapsto\left(f(t)\phi, \frac{P_{\phi}}{f(t)}+g(t){\phi}\right),
\ee
which includes a contribution to the new momentum which is linear in the field $\phi$.

Finally, let us mention that completely analogous results have been obtained for the remaining linearly polarized Gowdy models, namely those with spatial sections with topologies $S^1\times S^2$ and $S^3$. This analysis was performed in the following independent steps. First, the classical models were addressed in Ref. \cite{barbero1}, showing that, in these cases, the local gravitational DoF can also be parametrized by a single scalar field, namely an axisymmetric field in $S^2$. Then, following a procedure similar to the one introduced by Corichi, Cortez, and Mena Marug\'an, a Fock quantization with unitary dynamics was obtained \cite{barbero2}. In particular, a time-dependent scaling of the original field is again involved, now of the form $\phi\mapsto\sqrt{\sin t}\phi$. Finally, the uniqueness of the quantization obtained in this way was proved in Ref. \cite{cmvS2}. 

\subsection{Quantum Field Theory in Cosmological Settings}

A common feature of the Gowdy models mentioned in the previous section is that the local DoF are parametrized by a scalar field effectively living in a compact spatial manifold. Moreover, after the crucial scaling of the field, the dynamics is that of a linear field with time-dependent mass, i.e. it obeys a second order equation of the type
\begin{equation}
\label{sf-with-tdp}
\ddot{\chi}-\Delta~\chi +s(t)\chi=0,
\end{equation}
where $\Delta$ is the Laplace-Beltrami (LB) operator for the spatial sections in question, e.g. $S^1$ for the Gowdy model on $T^3$ and $S^2$ for the remaining two models. 

Remarkably, a whole different type of situations in cosmology can also be described by an equation of the form (\ref{sf-with-tdp}). Let us consider e.g. a free  scalar field in a homogeneous and isotropic FLRW spacetime, with line element
\begin{equation}
\label{FLRW-line-element}
ds^{2}=a^{2}(t)\left[-dt^{2}+{{h}_{ab}dx^{a}dx^{b}}\right],
\end{equation}
where $t$ is the conformal time, $a(t)$ is the scale factor, and ${h}_{ab}$ ($a,b=1,2,3$) is the Riemannian metric, for  either flat Euclidean space or the 3-sphere. A minimally coupled scalar field of mass $m$ obeys in this cosmological spacetime the equation 
\begin{equation}
\label{FLRW-eq-mot}
\ddot{\phi}+2\frac{\dot{a}}{a}\dot{\phi}-\Delta~\phi+m^{2}a^{2}\phi=0,
\end{equation}
where $\Delta$ is the LB operator defined by the metric ${h}_{ab}$ of the spatial sections. 

The most obvious situation described by this setup is the propagation of an actual (test) scalar matter field (disregarding the backreaction) in an FLRW background. Nonetheless, the treatment of quantum perturbations, both of matter and of gravitational DoF, also fits the above description.  In fact, in the context of cosmological perturbations, the leading-order approximation in the action, together with a neglected backreaction, amounts to keep the homogeneous classical cosmology as the background and treat both matter and gravitational perturbations as fields propagating on that background \cite{Mukhanov,Mukhanov-etal,Bardeen-PRD22,Halli-Haw-PRD31,fmmov}.

The quantum treatment of the  situations described above therefore faces the ambiguity of the choice of quantum \rep. In particular, the criterion based on stationarity, mentioned in Section \ref{weyl}, is not available, since all these situations are inherently non-stationary. 

There is however a natural avenue to address this issue, arriving hopefully at a unique quantum theory, with unitary dynamics. The way forward is actually suggested by a standard procedure, commonly found precisely in QFT in curved spacetimes (see e.g. Refs. \cite{BD,Fb}) and in the treatment of cosmological perturbations (see for example Refs. \cite{Mukhanov,Mukhanov-etal,Bardeen-PRD22,Halli-Haw-PRD31}). It consists in the scaling of the field variable $\phi$ by means of the scale factor, thus introducing the rescaled field $\chi=a(t)\phi$, which now obeys an equation of the type (\ref{sf-with-tdp}), with $s(t)=m^2 a^2- (\ddot a/a)$.
 
Thus, the combined effect of the use of conformal time and the scaling $\phi\mapsto a(t)\phi$ is to recast the field equation in the form (\ref{sf-with-tdp}), which is effectively the field equation of a linear field with a time-dependent quadratic potential $V(\chi)=s(t)\chi^2/2$, propagating in a static background with metric 
\be \label{29}
ds^{2}=-dt^{2}+{{h}_{ab}dx^{a}dx^{b}}.
\ee 
In this manner, we see that a generalization of the uniqueness result obtained for the Gowdy models would provide a useful criterion to select a unique quantization for fields in non-stationary backgrounds such as FLRW universes, typically with compact spatial sections with the topology of $S^3$ or $T^3$, allowing also applications to the usual flat universe case. Note that the restriction to spatial compactness is most convenient from the viewpoint of mathematical rigor, as otherwise infrared issues would plague the analysis. Nonetheless, the physical effects of the artificially imposed compactness, e.g. in the spatially flat case, should be irrelevant when the physical problem at hand does not involve arbitrarily larges scales, e.g. going beyond the Hubble radius. Moreover, in fact the case of continuous scales, which corresponds to non-compactness, can be reached in a suitable limit for flat universes, after completing all the demonstrations of uniqueness in the framework of compact spatial sections \cite{continuum}.

The desired generalization of the result about the uniqueness of the \rep\ in the aforementioned context of test fields and perturbations on cosmological backgrounds was obtained in Refs. \cite{cmsv,cmv2,cmov2,cmov4,flat}. In the rest of this Section, we very briefly explain the implications of these results.

Let $\mathbb{I}\times \Sigma$ be a globally hyperbolic spacetime, where $\mathbb{I}\in\R$ is an interval and $\Sigma$ is a~compact Riemannian manifold of dimension $d\leq 3$ (for cosmological applications, one can think of $\Sigma$ as being either $S^1$, $S^2$, $S^3$, or $T^3$). The spacetime is assumed to be static, with metric given by Equation (\ref{29}), where ${h}_{ab}$ is the time-independent, Riemannian metric on $\Sigma$. Consider a linear scalar field in $\mathbb{I}\times \Sigma$ obeying a field equation of the type (\ref{sf-with-tdp}), where $s(t)$ is (essentially\footnote{Only very mild technical conditions on $s(t)$ are required, see Ref. \cite{cmv2}.}) an arbitrary function and $\Delta$ is the LB operator associated with the metric ${h}_{ab}$. Note that any symmetry of this metric is transmitted to the LB operator, and therefore to the equations of motion. Let us consider the Weyl algebra associated with the field $\chi$ (and of course its canonical conjugate momentum) and their Fock \reps. Then,
\begin{enumerate}
\item{there exists a Fock \rep\ defined by a state which is invariant under the symmetries of the metric ${h}_{ab}$ (or equivalently, a Fock \rep\ with invariant vacuum) and such that the classical dynamics can be unitarily implemented;} 
\item{that \rep\ is unique, in the sense that any other Fock \rep\ defined by an invariant state and allowing a unitary implementation of the dynamics is unitarily equivalent to the previous one.}
\end{enumerate}

Remarkably, it again follows that the rescaling of the  field $\phi\mapsto a(t)\phi$ is quite rigid and uniquely determined: no other time-dependent scaling can lead to a(n invariant Fock) quantization with unitary dynamics, and so no ambiguity remains in the choice of the preferred field configuration variable. Furthermore, the unitarity requirement essentially selects as well the canonical momentum field. Physically, this can interpreted as  a unique splitting between the time-dependence assigned to the background and that corresponding to the evolution of the scalar field DoF.

Finally, it is worth mentioning that extensions of these uniqueness results where obtained in several directions. First, analogous results were obtained for scalar fields in homogeneous backgrounds of the Bianchi I type \cite{bianchiu}. We point out that these spacetimes are not isotropic, and therefore the conformal symmetry which was a common characteristic of the previous cases (at least asymptotically for large frequencies) is no longer present in an obvious way. Even more remarkable are the extensions of the uniqueness results attained for fermions, since they mark the transition to a largely unexplored territory \cite{ferm1,ferm2,ferm3}. A recent account of these results\footnote{Part of the techniques employed for fermions were already explored in the case of the scalar field in Bianchi I, in order to deal with the lack of conformal symmetry. A review of the range of different methods and improvements required to address the increasing degree of generalization encountered in the treatment of the scalar field can be found in Ref. \cite{Cortez:2019orm}.} can be found in Ref. \cite{rev_fermi}.

\section{Conclusions}

We have reviewed and discussed several results concerning the quantization of systems with relevant applications in cosmology. In particular, we have focused our attention on results ascertaining the uniqueness of the quantization process. This uniqueness is crucial in order to provide physical robustness to the eventual cosmological predictions of the quantum models in question, which would otherwise be fundamentally affected by ambiguities.

Starting with homogeneous models in this cosmological context, we have discussed a uniqueness result by Engle, Hanusch, and Thiemann, concerning the \rep\ of the Weyl relations commonly used in LQC. Turning to recent investigations carried out by our group and collaborators, the quantization of the family of inhomogeneous cosmologies known as the linearly polarized Gowdy models was reviewed next. Crucial in this discussion is the requirement of unitary implementation (at the quantum level) of the dynamics. Together with invariance under spatial symmetries, the criterion of unitary dynamics has proved very effective in the selection of a unique and physically meaningful quantization, in a variety of situations. A common general mathematical model embodying all these cases is that of a scalar field with a time-dependent mass (with arbitrary time dependence except for some very mild conditions), propagating in a static spacetime with compact spatial sections. Existence and uniqueness of a Fock quantization with unitary dynamics has been proved for such general model, thus providing unique quantizations in cosmological systems, ranging from quantum fields in FLRW backgrounds to the quantization of (perturbative) gravitational DoF.

In particular, it follows from our analysis and the aforementioned discussions that the quantization of linear transformations of physical interest via a unitary implementation does not necessarily require the existence of an invariant state, and that it is worthwhile pursuing alternatives that are not based on such invariance. 
In fact, what is really required in physical terms is a unitary implementation and not necessarily an invariant state, i.e.~unitary implementations via non-invariant states are still physically acceptable and cannot be discarded.
Clearly, a proof on the uniqueness of the quantization based just on the unitary implementation of those transformations, rather than on the existence of an invariant state, is a stronger result inasmuch as the requirement of unitarity is weaker than invariance.  Note that whereas in some circumstances there are good reasons to restrict attention to invariant states (for instance when there is a time-independent Hamiltonian giving rise to a group of transformations), 
in the general case of arbitrary transformations the requirement of an invariant state is not so compelling and non-invariant states cannot be simply disregarded. 
In this sense, it is worth pointing out that invariance can actually be a too much rigid demand in some situations, especially if the considered transformations provide a notion of evolution that is crucial to describe the dynamics. For instance, the unitary implementation of the evolution via (dynamical-)invariant states in typically non-stationary scenarios is of doubtful physical use, and in fact seems hopeless, whereas physically viable states, still leading to unitary dynamics and ensuring uniqueness, are nevertheless available and have direct cosmological applications. In this respect, we conclude with the following remark.

The uniqueness of the Fock quantization of the scalar field with time-dependent mass was obtained by restricting the attention to states that remain invariant under spatial symmetries. Although the removal of this restriction seems unlikely to lead to physically new \reps\ with unitary dynamics, it is nevertheless an open possibility. 
More precisely, it remains to be disproved the existence of representations with unitary
dynamics and a unitary implementation of the spatial symmetries which are nevertheless not unitarily equivalent to the representation defined by an invariant vacuum. Thus, a conceivable line of future research in this
area is to consider also Fock representations that, while
not possessing an invariant vacuum, still allow a unitary implementation of the spatial symmetries.
The uniqueness result would be strengthened if those representations were shown to be equivalent to the previous one, or otherwise new and potentially interesting representations could emerge.
Another possibility, with a clearly much higher potential to produce physically inequivalent results, is to use polymer inspired \reps\ for the scalar field, instead of Fock \reps.

\section*{Appendix:  Sketch of the Proof of Uniqueness of the Representation for the Scalar Field with Time-Dependent Mass in $S^1$}

In this Appendix, we are going to consider the simplest example of a linear scalar field in a static spacetime with compact spatial sections, namely the case where the spatial sections are 1-dimensional with the topology of the circle. The field equation reads
\be
\label{fes1}
\ddot{\chi}-\Delta{\chi}+s(t)\chi=0,
\ee
where the mass term $s(t)$ can be an (essentially) arbitrary function of time.

Since the spatial manifold is compact, and the field equation is linear, a Fourier decomposition gives us a discrete set of independent modes: 
\be
\label{fourier}
\chi(\theta,t)=\frac{1}{\sqrt{2\pi}}\sum_{n=-\infty}^{\infty}{\chi_n (t)}e^{i n\theta}=
\frac{1}{\sqrt{\pi}}\sum_{n=1}^{\infty}\bigl({q_n}\cos(n\theta)+{x_n}\sin(n\theta)\bigr)+\frac{q_0}{\sqrt{2\pi}}.
\ee
The configuration space for the scalar field is then described by the set of real variables $q_n$, $n\geq 0$, and $x_n$, $n>0$, which are completely decoupled. For simplicity, we drop all the modes $x_n$ (which can be treated like their cosine counterparts $q_n$ for $n>0$) and $q_0$, and continue with the infinite set $\{q_n, n>0\}$. The variable $q_0$ is dropped just to avoid introducing a special treatment in the case $s(t)= 0$, since $n=0$ corresponds to a zero frequency oscillator, or a free particle, instead of a regular harmonic oscillator. In any case, it describes a single degree of freedom, which cannot affect the considered matters of unitary implementation.

The equations of motion for the modes are
\be
\label{eqmm}
{\ddot q}_n+[n^2+s(t)]q_n=0.
\ee
The corresponding Hamiltonian equations are
\be
\label{ham}
\dot q_n=p_n,\quad \dot p_n=-[n^2+s(t)] q_n,
\ee
where $p_n$ is the momentum canonically conjugate to $q_n$, i.e.~$\{q_n,p_{n'}\}=\delta_{nn'}$.

There are many \reps\ of the CCRs satisfied by the infinite set of pairs $\{(q_n,p_n), n>0\}$, or of the associated Weyl algebra. For instance, every sequence $\{\mu_n, n>0\}$ of (quasi-invariant\footnote{In order to provide a unitary \rep\ of translations, measures are required to satisfy the technical condition of quasi-invariance, which is satisfied e.g.~by any Gaussian measure.}) {\em probability} measures in $\R$ gives a \rep, since it defines a regular product measure in the set of all sequences $(q_1,q_2,\ldots)$, thus providing a Schr\"odinger type of quantization, in the Hilbert space of square integrable functions in the configuration space. What is not available, however, is the straightforward generalization of the usual \rep\ in finite dimensions obtained from the Lebesgue measure $dq$, since no mathematical sense can be made of the formal infinite product $\prod_{n=1}^{\infty}dq_n$. In such context, Fock \reps\ of the Weyl relations are given by (normalized) Gaussian measures, which still make perfect sense in infinite dimensions. Even after restricting our attention to Gaussian measures, there are endless possibilities, and many of them lead to inequivalent \reps.

In the present case, a particularly important measure in our infinite dimensional configuration space is
\be
d\mu=\prod_{n=1}^{\infty}e^{-n\,q_n^2}\, \frac{\sqrt{n}dq_n}{\sqrt{\pi}},
\ee
which is associated with the quantum operators
\be
\hat q_n \Psi= q_n\Psi, \quad \quad \hat p_n\Psi= -i \frac{\partial}{\partial q_n}\Psi
+i n q_n\Psi.
\ee
This is in fact  a particular realization of the Fock \rep\ given by the complex structure $J_0$, defined as in Eq. (\ref{jm})\footnote{With the obvious adaptations, taking into account that the spatial manifold is now $S^1$ instead of $\R^3$.} with $m$ equal to zero. So, it should be no surprise that this particular representation allows a unitary implementation of the free massless field dynamics (i.e. with $s(t) = 0$). The quantization is the most natural one for the massless field, since the wave functional $\Psi = 1$ is invariant under the unitary group $U(t)$ implementing the dynamics. Putting it differently, there is a well defined quantum Hamiltonian, and $\Psi = 1$ is the zero-energy state of the free massles field.

The same representation also allows a unitary implementation of the time-dependent mass case. To see this, let us introduce the annihilation and creation-like variables $a_n$ and ${\bar a}_n$, defined by
\be 
{a_n}=\frac{nq_n+ip_n}{\sqrt{2n} } ,
\ee
which are precisely the ones associated with the complex structure $J_0$. In particular, this means that $J_0$ takes the diagonal matrix form ${\rm diag} (i,-i)$ when written in terms of the basis in phase space made of the pairs $(a_n, {\bar a}_n)$. With respect to this basis of (complex) variables, the classical evolution (from time $t_0$ to time $t$) determined by Equation (\ref{ham}) is given by non-vanishing $2\times 2$ matrix blocks of the form
\be
\label{beta}
{\cal U}_n(t,t_0)=
\left(\begin{array}{cc} \alpha_n(t,t_0) & \beta_n(t,t_0)\\ {\bar \beta}_n(t,t_0) &
{\bar \alpha}_n(t,t_0) \end{array}\right),
\ee
with
\be
|\alpha_n(t,t_0)|^2-|\beta_n(t,t_0)|^2=1,\ \ \qquad \forall n>0, \ \forall  t,t_0.
\ee
Note that, in the case $s(t)= 0$, precisely because of the way in which the variables $a_n$ are defined, the above parameters $\beta_n$ are all vanishing (and the parameters $\alpha_n$ are just phases). 

For quite general functions $s(t)$ in Equation (\ref{ham}), it turns out that the following condition is satisfied \cite{cmsv}:
\be
\label{sqs}
\sum_n^\infty|\beta_n(t,t_0)|^2<\infty, \quad \forall t,t_0.
\ee
This is precisely the necessary and sufficient conditions for unitary implementability of the dynamics, in the $J_0$ \rep. Thus, there is a Fock \rep, defined by a complex structure that remains invariant under the spatial symmetries, such that a unitary quantum dynamics can be achieved.

The proof that a quantization with the above characteristics is unique goes as follows. To begin with, any other invariant (under spatial isometries) complex structure $J$ is related to $J_0$ by $J=KJ_0K^{-1}$, where $K$ is a symplectic transformation given by a block diagonal matrix with $2\times 2$ blocks of the form
\be
\left(\begin{array}{cc}\kappa_n & \lambda_n \\ {\bar \lambda}_n &
{\bar \kappa}_n
\end{array}
\right),
\qquad|\kappa_n|^2-|\lambda_n|^2=1, \ \ \forall n>0.
\ee
Suppose  now that  the dynamics is unitary in the Fock \rep\ defined by $J$. It turns out that this is equivalent to the unitary implementability in the $J_0$-\rep\ of a modified dynamics, obtained precisely by applying the transformation $K$ to the canonical transformations corresponding to time evolution. A simple computation shows that, for this modified dynamics, the coefficients $\beta_n$ in Equation (\ref{beta}) are replaced with
\be
\beta^J_n(t,t_0)=2 i {\bar \kappa}_n\lambda_n
{\it Im}[\alpha_n(t,t_0)]+({\bar \kappa}_n)^2\beta_n(t,t_0)-\lambda_n^2{\bar \beta}_n(t,t_0),
\ee
where the notation ${\it Im}$ denotes the imaginary part. Then it follows from the hypothesis of unitary dynamics that the following condition holds:
\be
\label{sqsJ}
\sum_n^\infty|\beta^J_n(t,t_0)|^2<\infty,\qquad \forall t,t_0.
\ee
Now, a detailed asymptotic analysis \cite{cmsv,cmv2}  shows that condition (\ref{sqsJ}) implies that
\be
\sum_n^\infty|\lambda_n|^2<\infty.
\ee
Given the relation between $J$ and $J_0$, this last condition guarantees precisely that the operator  $J-J_0$ is of the Hilbert--Schmidt type, i.e. that the Fock \reps\ defined by $J$ and $J_0$ are unitarily equivalent.

\vspace{6pt}

\acknowledgments
The authors are grateful to B. Elizaga Navascu\'es for discussions. This work was supported 
by the Spanish MINECO grant number FIS2017-86497-C2-2-P, the Spanish MICINN grant number PD2020-118159GB-C41, and the European COST (European Cooperation in Science and Technology) Action number CA16104 GWverse.
J.M.V. is grateful for the support given by research unit Fiber Materials and Environmental Technologies (FibEnTech-UBI), on the extent of the project reference UIDB/00195/2020, funded by the Fundação para a Ciência e a Tecnologia (FCT).


\newpage
	

\end{document}